\documentstyle[eqsecnum,aps,twocolumn,psfig]{revtex}

\begin{document}
\draft
\title{Line shape narrowing in carbon dioxide at high pressures }
\author{Vladimir F. Golovko}
\address{Laboratory of Theoretical Spectroscopy, Institute of Atmospheric Optics, SB RAS, \\
1, Akademicheskii Av., Tomsk 634055, Russia }
\maketitle

\begin{abstract}
The description of the narrowing effect at high pressures is demonstrated on
the example of CO$_{2}$-He absorption for Q-, P-, and R-branches and for the
head of the R-branch. The problem is focused on the physical meaning of this
phenomenon that should be associated with the angular and translation
momentum transfer to matter from the photon absorbed. The effect is
described in the approximation of the single line without the conventional
line mixing. A narrowing function is introduced that links the absorption in
the resonance region with the one in far wings. This function enlarges the
absorption of a single line approximately in two times in the resonance
region at high pressures and decreases the one approximately in 20 times in
far wings in comparison with the Lorentzian, independently of the type of
molecule and pressures. The halfwidth of a single line is assumed to have a
nonlinear pressure dependence and approaches to a saturated value at high
pressures.
\end{abstract}


\preprint{HEP/123-qed}



\section{INTRODUCTION}

The present paper is devoted to the line shape narrowing or the so-called
''line mixing effect'' non less puzzling phenomenon of line shapes theories
than the intriguing problem of extreme wings [1,2]. The narrowing of the
absorption line shape for high enough pressures leads to the evident
super-Lorentzian absorption near band centers and this effect had been first
reported by Bloembergen, Purcell, and Pound [3] more than 50 years ago. At
present, this phenomenon is known for many molecules, bands, in many
spectral regions and has many applications (see, e.g., reviews by L\'{e}vy
et al. [4] and by Sp\"{a}nkuch [5]). Near the band head, this effect had
been observed by Grigor'ev et al. [6] for the band 3$\nu _{3}$ of CO$_{2}$.
The recent experiments and modelings of the line mixing for various bands
and branches may be found in Refs. [7-10]. The above-mentioned papers [6-10]
as well as ideas, experimental facts, and observations from Refs. [11, 12]
are directly used in the present studies. The simplicity of computations
that has underlined the importance of application of the line mixing in
atmospheric studies had been pronounced in Refs. [11, 13, 14].

This paper continues our line shape exploration efforts devoted to
investigations of CO$_{2}$ absorption in far wings and main results had been
first announced on the Symposium IRS-2000 [15]. These studies are based on
new theoretical ideas that are nearly to Ref. [16] and especially to Ref.
[17] and they are aimed to characterize the absorption line shape in
resonance and far wing regions by some common principles. The problem
numerates many approaches for its decision with many ideas and we do not
pretend to accomplish a comprehensive study of this question and even to
outline some appreciable part of the pertinent works in these regions.

\section{GENERAL PRINCIPLES AND THE LINE SHAPE}

There is a conventionally established opinion that the Lorentz's contour
cannot describe actual molecular spectra. However, it seems that the problem
is not so lucid. The matter comprises molecules that, in turn, are
associated with many cavities in the form of two-level systems and any
photon propagated within the matter feels these cavities, i.e., there is a
probability that this photon will be captured by any of them. The latter
means that any photon being in the vicinity of a molecule of interest with
any frequency has a probability to interact with any cavity of this
molecule, independently of the tuning wavenumber.

As a rule, the dispersion (Lorentzian) line shape was used to derive
description of this interaction by methods of the wave mechanics. In fact,
no contour other than the Lorentzian had been employed in the overwhelming
majority of theories. For example, even the so-called Rozenkranz's line
shape [18] that is a basic approach for many sophisticated theories and
techniques is not something else as the modified dispersion profile. Our
resent results can encourage this idea, since the Lorentzian priority may
cause the original description of the sub-Lorentzian properties of CO$_{2}$
absorption [15]. The exclusion may be the statistical line shape that uses
ideas of the quantum theory, where the photon energy emitted or absorbed
corresponds to the difference between two energy levels being in the
statistically changing field of other particles [19]. However, the
statistical line shape that is fruitful for description of the line wing
broadening in plasmas, nevertheless, does not exclude the Lorentzian in the
core part [20] as well.

The photon interaction with a two-level system can lead to the photon
absorption, if the momentum and angular momentum have been conserved.
However, the photon cannot interact with a molecule or molecules infinitely,
because there is a need of time to transfer the momentum and angular
momentum from photon to the molecule, because every molecule has the inertia
that is determined by the mass or the inertia momentum. This means that an
act of absorption or emission begins, if the system is in a determined
stationary state not in an intermediate position, and some time is needed in
order the system due to its inertia can change its stationary states. This
interpretation completely corresponds to the main principles of the quantum
mechanics. A period of the angular and translation momentum transfer
restricts the duration of the absorption by some minimal value, therefore
the lifetime of the exited state to reckon in the conventional approaches is
also not less than some minimal value.

The relaxation time $\tau _{rot}$ of the excited rotational state $i$ into a
lower state $j$ or the lifetime of the excited rotational state $i$ in
presence of the lower state $j$ is on average determined as 
\begin{equation}
\tau _{rot}^{ij}=1/2c\gamma _{rot}^{ij},
\end{equation}
where $c$ is the light velocity (cm/c) and $\gamma _{rot}^{ij}$ is the
halfwidth (cm$^{-1}$) of the rotational line $ij$. Let us omit the
superscripts in Eq. (2.1) in further. On average, the rotational angular
momentum transfer time cannot be shorter than some limits that are
determined as reciprocal separation between rotational lines, 
\begin{equation}
\tau _{rot}\geq \tau _{rot}^{\min }=1/2c\gamma _{rot}^{s},
\end{equation}
where the saturation halfwidth $\gamma _{rot}^{s}\sim \overline{\Delta
\omega _{rot}}$, and $\overline{\Delta \omega _{rot}}$ is this mean
separation. The similar relationships must be for the translation momentum
transfer time 
\begin{equation}
\tau _{vib}\geq \tau _{vib}^{\min }=1/2c\gamma _{vib}^{\max }\sim 1/c\Delta
\omega _{vib},
\end{equation}
where $\gamma _{vib}$ and $\Delta \omega _{vib}$ should be regarded as the
halfwidth of the selected vibrational band and the separation between bands,
respectively. Our assumption consists of that the minimal duration of the
rotational or vibrational relaxation is related to the conservation laws of
the angular or translation momentum, respectively.

The finitely collision duration hypothesis [19, 21] is distinguished from
our assumption. It was used to be fruitful for explanation of the line
mixing [22] in CO$_{2}$ and the wing absorption in H$_{2}$O [23]. However,
this idea may find objections from the point of view of collision and
statistical theories [19, 20]. The main reason is that the impact
approximation (infinitesimal collisions) describes good the line shape
nearly resonances, where transitions are caused by slow molecules, i.e.,
with finitely collision time, but this approximation is lame in wings that
must be reasoned by rapid collisions, i.e., with the infinitesimal collision
time. While according to the finitely collision duration, the effect could
be inverse. Though these objections have a paradoxical character, indeed,
they are serious because the collision duration mostly relies on the
relative velocity of molecules not their internal properties.

The paradox may be removed by the fact that the duration of the photon
interaction differs from the collision time and the hypothesis in our
edition means that there is a minimum limit of time of the photon
interaction with a molecule. The angular or translation momentum transfer
from photon to the molecule or vice versa must be occurred during this time
that because of the molecule inertia cannot on average be less than the
minimum limit.

Note that for the determination of the separation between lines or bands in
CO$_{2}$ molecule, the forbidden transitions must not be formally taken into
account. These mean they do not change the number of channels that could be
appropriate for the momentum transfer. Certainly, the pressure-induced
absorption is here ignored. The appearance of the supplement resolved lines
(e.g., rotational lines in Q- branches of the perpendicular bands in CO$_{2}$%
) enlarges the transfer time minimal limit under the channels of interest.
The supplement lines effect the narrowing of the line shape that formally
approves the name of the ''line mixing'' in Q-branches. Comparing the
different branch, thereby one may conclude that weak lines between strong
lines decrease the transfer time limit and strong lines increase this limit.
The transfer time limit determined by the far lines always should not be
greater than those done by adjacent lines. The role of the adjacent lines
always should be more significant than ''interaction'' between far lines
and, as one can see in Section V, the adjacent lines mostly determined the
effect of narrowing of the lines in bands at high pressures.

\section{PRESSURE-DEPENDENT LINE HALFWIDTH}

There is a well established empirical fact that the line shape of a single
line is described by the Voigt contour at any pressures of interest and
adjustments of the line parameters stored in databases are practically
carried out by this profile. The collision halfwidth of a single line
strongly depends on pressure (Fig. 1). The database contains the values of $%
\gamma _{atm}$ that equals to $\gamma \left( p\right) -\gamma _{0}$ (Fig. 1)
at $p$ equal to 1 atm. The halfwiths at other pressures $\gamma \left(
p\right) $ is as a rule determined by the linear law including high
pressures, where experimental evaluation is difficult. 

The part of the curve for small and moderate pressure less than, e.g., $%
p_{L} $ is presented in Fig. 1, following Ref. [24]. We believe that the
Lorentzian should be a basic contour of the line shape generated in this
paper. Let us assume that the halfwidth of the Lorentzian at high pressures
is deviated from being linear on pressure (Fig. 1) and approaching to the
saturation value not being greater than some value proportional to the
parameter $\gamma _{rot}^{s}$ (2.2). Therefore, it reaches the saturation at
some critical pressure $p_{s}$ as one sees in Fig. 1. Because of line
overlapping, it is impossible to measure this halfwidth at high pressures
and, thereby, to extrapolate the Voigt line shape on high pressures. In
order to verify this hypothesis, two estimations of the CO$_{2}$ absorption
in mixtures with helium have been made at the critical density $p_{s}=135.8$
atm (Fig. 2a) and $p=657.1$ atm (Fig. 2b). The Lorentzian halfwidth in these
computations (Figs. 2a and 2b) was the same. The latter does not concern the
far wings used for absorption anticipation in the inter-band region. The
description of the model and calculation details will be considered in
Sections IV and V, respectively.
\begin{figure}
\centerline{\psfig{figure=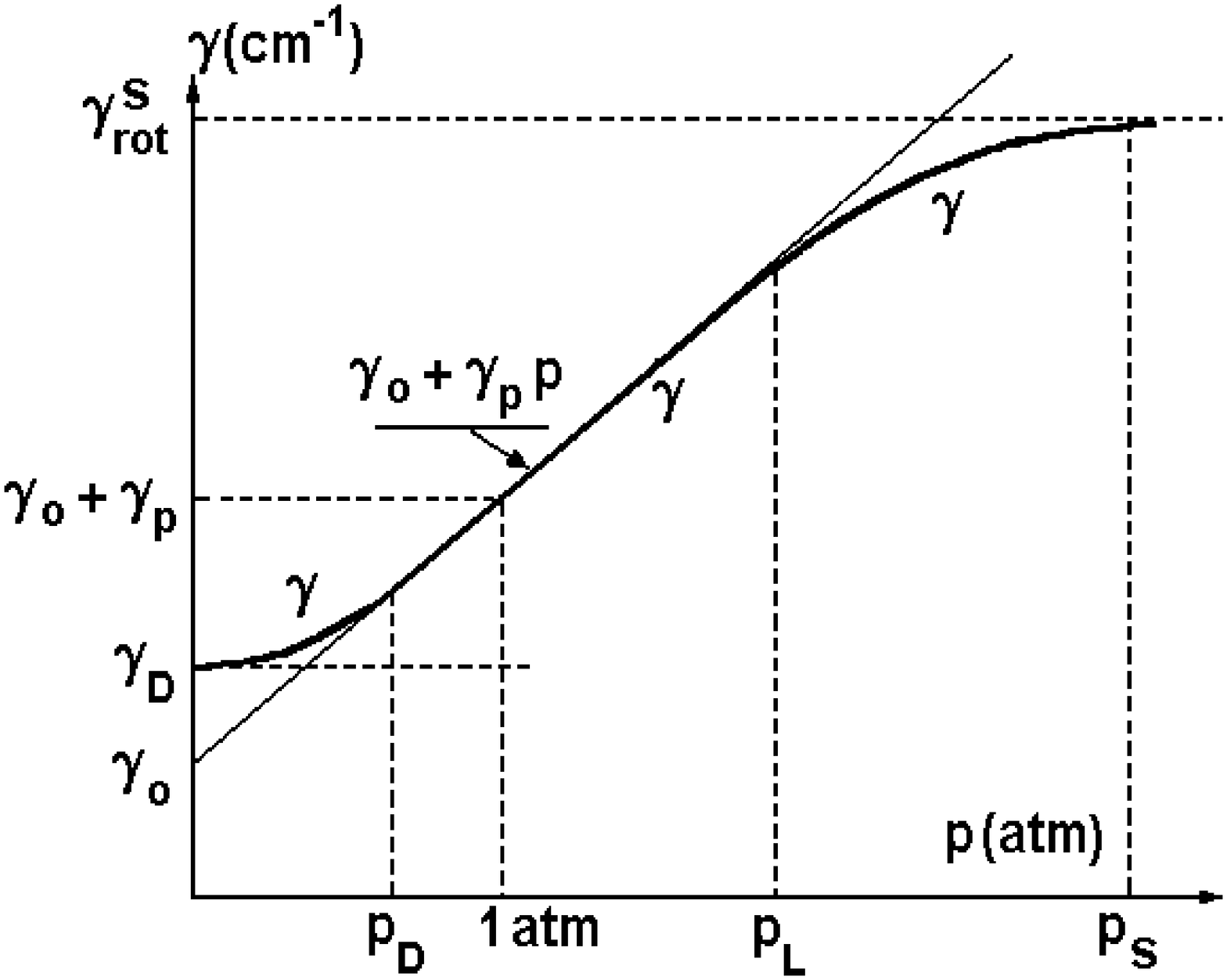,width=8cm,height=5.3cm,angle=270}}
\caption{A hypothetical behavior of the pressure-dependent collision
halfwidth of a single line demonstrates the main hypothesis of existence of
the minimum time limit of the photon interaction with a molecule. The
definition of the saturated value $\gamma _{rot}^{s}$ follows from Eq. (2.2)
as the separation between rotational lines, that approximately becomes equal
to this value at some critical pressure $p_{s}$, when the rotational
structure of spectrum is not noticeable. The pressure interval [$p_{D},p_{L}$%
] indicates the quasi-linear region that is mostly applied to the
experimental measurements of the collision halfwidth.}
\end{figure}

\section{NARROWING FUNCTION}

Interesting observations may be found in a few papers [11,12] (see other
cites in Ref. [12]). The absorption peak of the bands studied by the
Loretzian method is underestimated in two times [12] at those minimal
pressures, when the fine rotational structure of the bands becomes
unnoticeable. The absorption in near wings is overestimated also in two
times independently of pressures, Q-branch line spacing, and wing
wavenumbers [11,12]. These facts tell us that the narrowing effect is weakly
dependent on molecular properties similar to the exponential behavior
revealed for CO$_{2}$ far wings absorption (a few examples of these
computations are reported in Ref. [15]).

Applying the methods of probes and errors, we found out that the dispersion
line shape of the line $i$ has to be multiplied by some factor-function $%
\Gamma _{i}\left( \omega ,p\right) $ and by the exponential factor $%
\overline{n_{i}}\left( \omega \right) $ for $\omega \geq \omega _{i}$ or by $%
1-\overline{n_{i}}\left( \omega \right) $ for $\omega <\omega _{i}$, where $%
\omega _{i}$ is the resonance wavenumber (frequency) of the line $i$. The
exponential factor is not essential near the resonance spectral region and
it is not considered in details in this communication. In turn, let the
narrowing function be responsible for redistribution of the intensity in the
line shape, that may be associated with the momentum transfer. This
narrowing factor-function $\Gamma _{i}\left( \omega ,p\right) $ at pressures 
$p\geq p_{s}$ (Fig. 1) has the form 
\begin{equation}
\Gamma _{i}\left( \omega ,p\right) _{p\geq p_{s}}=\left( 1/4\right) ^{x_{i}},
\end{equation}
where the power $x$ is determined as follows: 
\[
x_{i}=x_{\min }\text{, if}\left| \omega -\omega _{i}\right| \leq a\gamma ; 
\]
\[
x_{i}=x_{\max }\text{, if}\left| \omega -\omega _{i}\right| \geq b\gamma ; 
\]
\begin{equation}
x_{i}=\frac{\left| \omega -\omega _{i}\right| -c\gamma }{b-c\gamma }x_{\max
},\text{ if }b>\left| \omega -\omega _{i}\right| >c\gamma \text{ ; }
\end{equation}
\begin{equation}
x_{i}=\frac{c\gamma -\left| \omega -\omega _{i}\right| }{c\gamma -a}x_{\min
},\text{ if }a<\left| \omega -\omega _{i}\right| \leq c\gamma \text{ .}
\end{equation}
\begin{figure}
\centerline{\psfig{figure=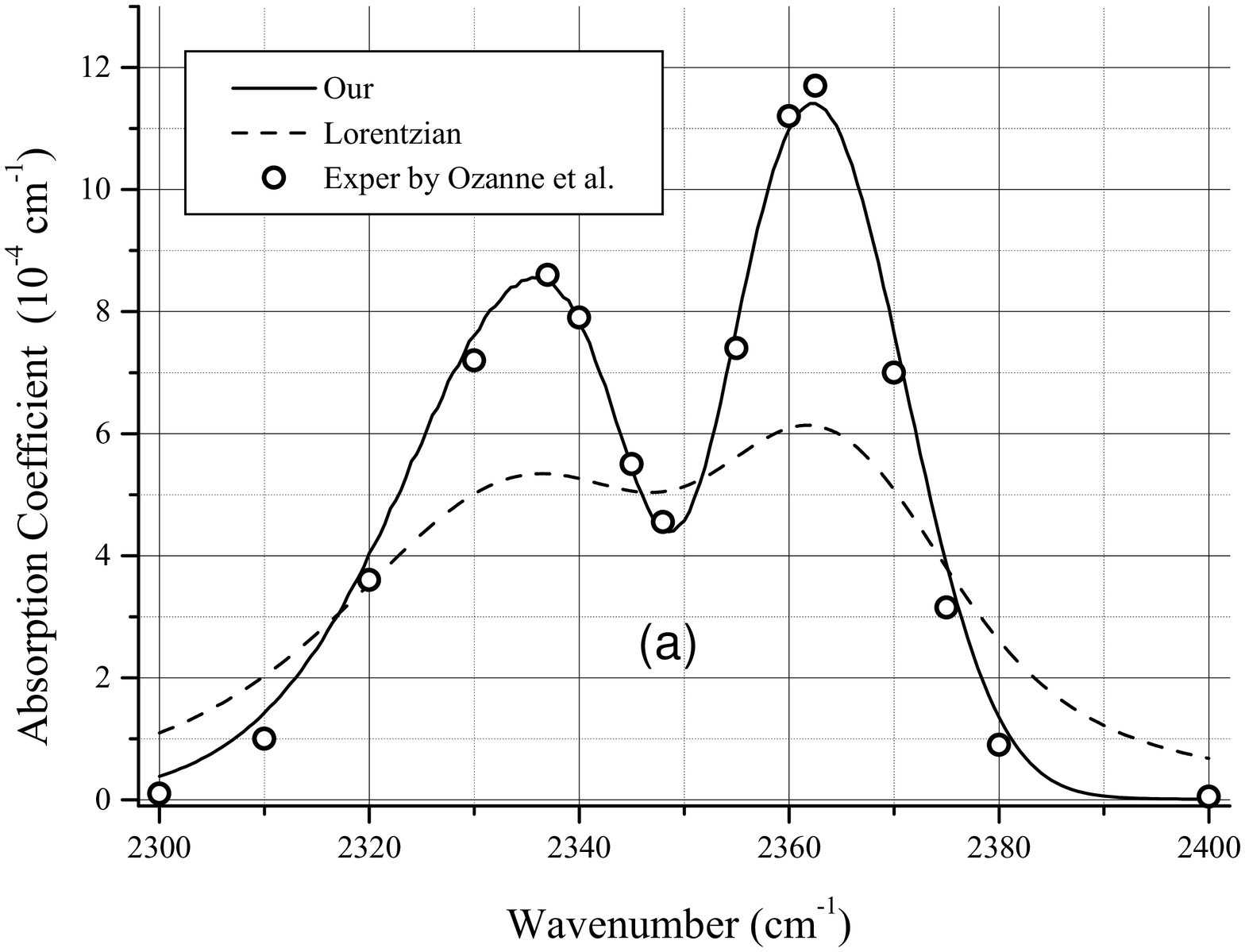,width=8cm,height=5.3cm,angle=270}}
\centerline{\psfig{figure=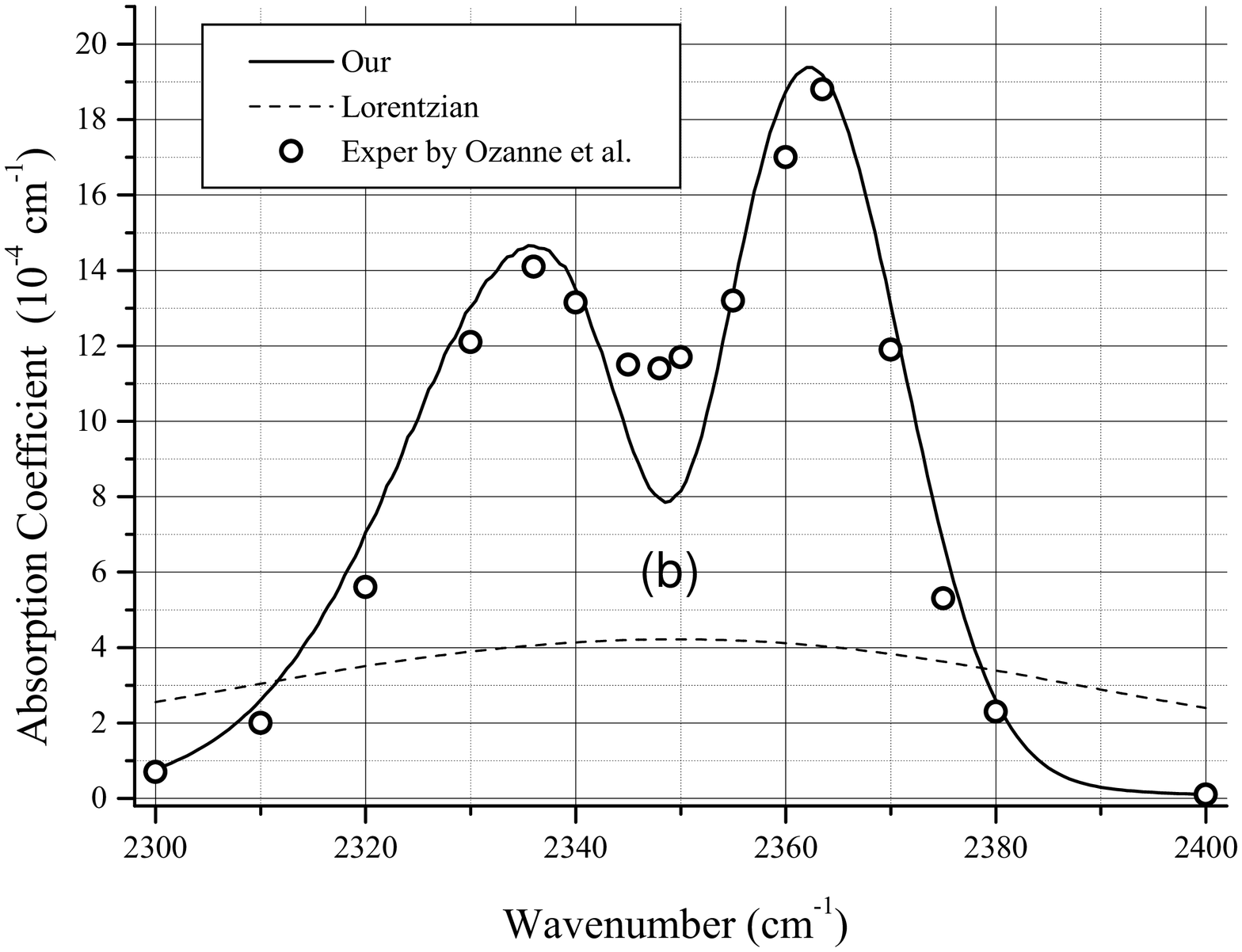,width=8cm,height=5.3cm,angle=270}}
\caption{Verification of the main hypothesis of the minimum time limit of
the angular momentum transfer on the example of the two calculations of the
absorption coefficient in the central region of the band $\nu _{3}$ near 4.3 
$\mu $m in mixtures with helium. Densities: (a) $n_{CO2}=1.63.10^{-5}$ Am
and $n_{He}=124.3$ Am (135.8 atm); (b) $n_{CO2}=2.73.10^{-5}$ Am and $%
n_{He}=603.4$ Am (657.1 atm). Experiment by Ozanne et al. [9] was performed
at the temperature 298 K.}
\end{figure}

The dimensionless parameters $a$, $b$, and $c$ are shown in Fig. 3. The
ordinary Lorentzian halfwidth $\gamma $ in Eqs. (4.1-3) at pressures ($p\geq
p_{s}$) is equal to the constant $\gamma _{rot}^{s}$. Then, $\gamma =\gamma
_{rot}^{s}$, at these pressures. The parameters $x_{\max }$ and $x_{\min }$
were studied by method of probes and errors near some values, so that, the
following equations would be valid 
\[
\Gamma ^{\max }\left( \omega ,p\right) _{p\geq p_{s}}=\left( 1/4\right)
^{x_{\min }}=4, 
\]
and 
\[
\Gamma ^{\min }\left( \omega ,p\right) _{p\geq p_{s}}=\left( 1/4\right)
^{x_{\max }}=0.1. 
\]
Thus, $x_{\min }=-1$ and $x_{\max }=\ln 0.1/\ln 0.25$.

\begin{figure}
\centerline{\psfig{figure=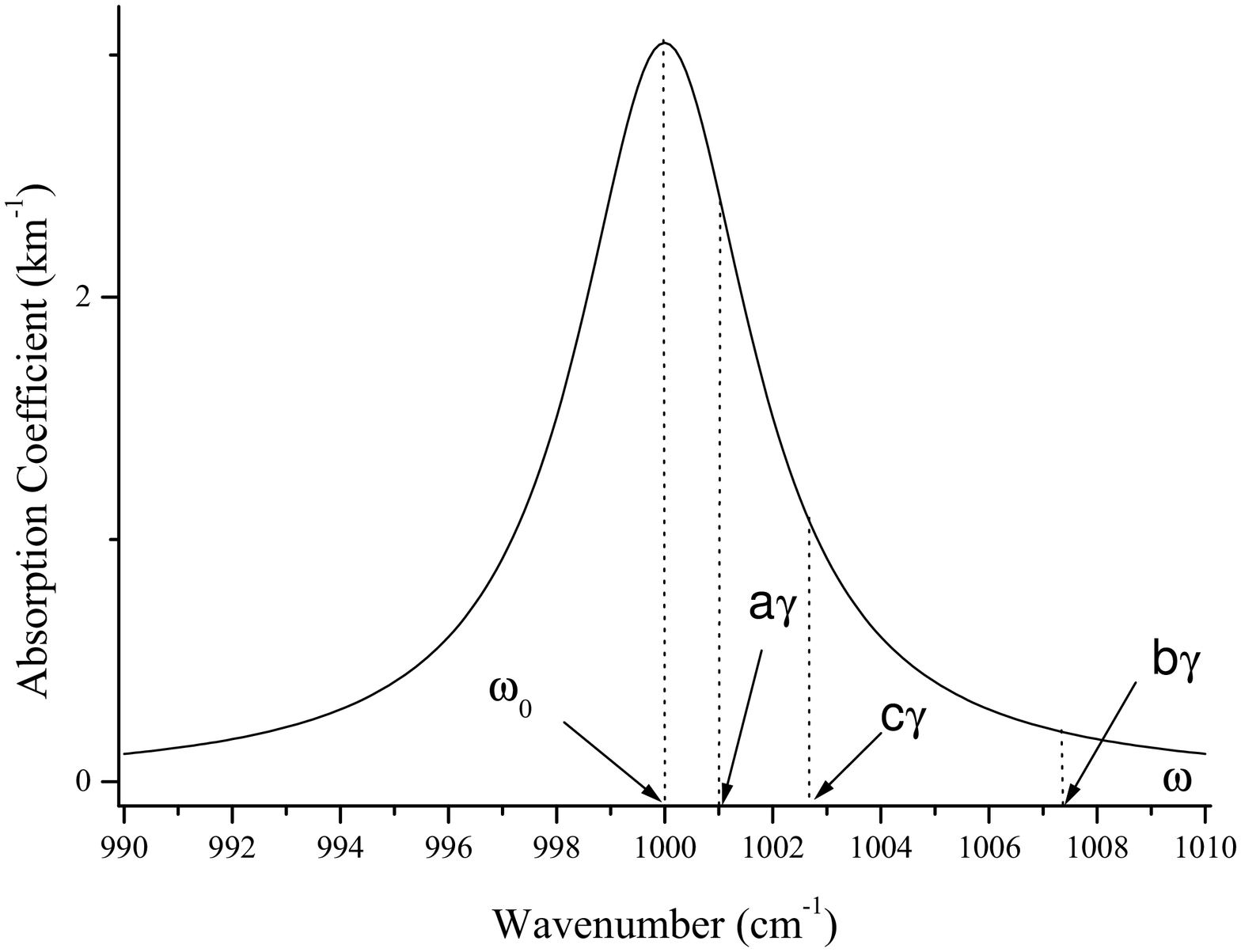,width=8cm,height=5.3cm,angle=270}}
\caption{The line shape regions for constructing the narrowing function $%
\Gamma _{i}\left( \omega ,p\right) _{p\geq p_{s}}$ (4.1).}
\end{figure}
The factor 2 at the factor-function $\overline{n_{i}}\left( \omega \right) $
is included in the function (4.1). Without this factor, the function (4.1)
has a simple assignment. In comparison with the Lorentzian line shape with
the saturation halfwidth $\gamma _{rot}^{s}$, it increases the intensity by
factor 2 in the region less than the parameter $a$ (Fig. 3) and, in
comparison with the Lorentzian line shape having the ordinary halfwidth, it
decreases the absorption in wings by the factor 0.05 for the wavenumber more
than parameter $b$ (Fig. 3). The functions (4.2) and (4.3) are sewing
together the factors 2 and 0.05. The calculations of the spectrum at
pressures $p\geq p_{s}$ are fulfilled along with the Lorentzian having the
saturation halfwidth $\gamma _{rot}^{s}$, therefore the parameter $x_{\max }$
is determined in the computational algorithm as 
\[
x_{\max }=\ln \left( 0.1\gamma /\gamma _{rot}^{s}\right) /\ln 0.25, 
\]
where the $\gamma $, as above, is the conventional Lorentzian halfwidth
linear to the pressure. In all our computations, e.g., in Fig. 2, the
relation $\gamma _{rot}^{s}=3.92\overline{\Delta \omega _{rot}}$ is used.
The separations $\overline{\Delta \omega _{rot}}$ are different for
different types of band branches.

The parameter values $a$, $b$, and $c$ for estimations in Fig. 2 are
specified as 0.72, 3.92, and 1.2. The discrepancies in Fig. 2b between data
calculated and measured may be explained by existence of the inter-branch
narrowing effect found out in paper by Qzanne et al. [9]. Following ideas of
the present paper, one may foresee that this effect for the photon
interaction with matter should be elucidated by existence of the minimum
time limit of the translation momentum transfer. Its rates may be different
for molecules in states with different angular momenta. Somewhat other
behavior for the branch interactions are modelled for the $3\nu _{3}$ band
and we will come back to this question in Section V.C.

The narrowing function may be assumed as the function controlled the
probability (rate) of the transition in a two-level system and that should
be independent on the intermolecular interactions. The latter is not
unexpected. The resonance nonlinear interaction of radiation with matter
that produces the transition probability oscillating with the Rabi frequency
[25] also does not depend on the intermolecular forces. According to this
resonance approximation, the probability in far wings must be also
oscillated, but with the detuning frequency [25]. Moreover, the classical
impact theories of the line shape by Lorentz, Lenz, and Weisskopf [19] are
based on the phase interruption due to strong impacts between molecules,
that destroys the emission in a two-level system. The existence of this
long-term emission was not clear in the classical theories, but it may be
replaced in a modern representation by the oscillating probability that is
caused by altering of level populations with the Rabi frequencies. In this
interpretation, the emission and absorption may be considered by the similar
technique.

Two different mechanisms of the angular momentum transfer have to be
accounted for the central region and wing of the line shape, respectively.
Let us interpret these mechanisms by a hypothesis of the paper [17]. Let two
coherent photons [17] interacts with a molecule as a fluctuation of four
[17] or five photons. The latter was used for calculations in Ref. [15]. Let
the angular momentum transfer occurs during the rotation of the polarization
plane of one coherent photon on right angle in respect to the second
coherent photon. Thus, two coherent photons or the whole fluctuation from
four-five photons partake in the angular momentum transfer. The absorption
rate (probability) being normalized in any reasonable theory per one photon
also increases in two times.

Another mechanism should be assumed in wings. Perhaps, the act of the direct
photon-molecule angular momentum exchange is impossible for big detuning
frequencies for a short period. The angular momentum transfer in wing
regions is delayed via an act of the photon-photon angular momentum
exchange, i.e., between two fluctuations. The rate (probability) normalized
per one photon should be less for the absorption act in $\left( 1/4\right)
^{2}$-$\left( 1/5\right) ^{2}$ times. We have found this value as equal to
1/20.

Thus, the physical meaning of the puzzling narrowing effect should be
associated, first of all, with the quantum representations in the form of
the photon fluctuations. The latter is also confirmed by the absorption
modeling in wings [15]. The line mixing method based on the Fano's approach
of the relaxation matrix [26] was in general successful in the line shape
theory, but its physical meaning should not be obligatory concluded as the 
interference of lines. Exact calculations of cross-relaxation constants 
in CO$_{2}$ by Petrova et al. [27] showed that they about in 
two-five orders of magnitude less than diagonal ones and cannot 
explain the narrowing effect at proper moderate pressures of N$_{2}$. 
Supplement hypotheses as, e.g., the statistical representation [18] 
must be incorporated and, then, the {\it ad hoc} and semiempirical 
parameters would have been included for practical calculations as it has 
been made in many works.

The narrowing is closer to definition of the rotational and vibrational
relaxation. However, the technique of this relaxation must consider
collisions that are occurred during the momentum transfer, i.e., for
transitions of active molecules from one stationary state to another. This
can be on average observed at high pressures. Since the quantum mechanics
admits consideration only of observable quantities assigned for states with
orthogonal functions, the problem becomes serious. Whether the apparatus of
quantum mechanics, e.g., in the form of the density matrix, suffices to
describe the effect of the collisions for the intermediate states? How means
the possible photon fluctuations related to a two-level system can be 
incorporated in the theory?

\section{VALIDATING THE NARROWING FUNCTION}

In fact, the narrowing function together with the exponential form of the
line shape reveals a good coincidence of the calculated absorption with the
experimental data in the wing region [15]. In order to verify the narrowing
function in the central region, the absorption coefficients are regarded for
the Q-, P-, R-branches of CO$_{2}$-He mixtures. Our modelings of absorption
in far wings show [15] the mixture of CO$_{2}$ with helium may be assumed
with weak nonlinear absorption [16] and then it is convenient for validating
the narrowing function hypothesis. The exponential form as well has the
lesser impact in the central region.

\subsection{$\nu _{3}$ band region}

Such calculations for the $\nu _{3}$ band are depicted in Fig. 2. The
narrowing function acts on the Lorentzian with the halfwidth $\gamma
_{He}=0.52\gamma _{N_{2}}$. This value is the same as in our calculations
for the mixture of CO$_{2}$ with helium in far wings [15] and it is nearly
to the experimental value evaluated in Ref. [2]. The saturated value of the
halfwidth $\gamma _{rot}^{s}$ is equal to 3.919$\overline{\Delta \omega
_{rot}}$, where for the $\nu _{3}$ region the $\overline{\Delta \omega _{rot}%
}$ is specified as 1.2 cm$^{-1}$ that approximately is the mean separation
between rotation lines in the R-branch. The parameter $b$ (Fig. 3) is 3.92.

On the whole, a good coincidence of data calculated and experimental is
observed (Fig. 2a), but the same predicted for the higher pressure in Fig.
2b is worse. These computations are performed with the same saturated
halfwidth $\gamma _{rot}^{s}$ for all pressures greater than some critical
pressure $p_{s}$ equal 135.8 atm. Contrary to the latter, the absorption
estimation in far wings is with the narrowing function that comprises 1/20$%
^{th}$ from the Lorentzian contribution with the conventional halfwidth
linear to pressure. The value of the latter is also 0.52$\gamma _{N_{2}}$.
The worse coincidence of the two compared plots in the center of Fig. 2b
should be associated with the inter-branch mixing effect introduced in Ref.
[9]. In our interpretation the latter should arise due to a lower time limit
(2.3) of the translation momentum transfer to active molecules from the
photon absorbed.

\subsection{Q-branch of the $\nu _{2}$ band region}

Thus, the halfwidth $\gamma _{He}=0.52\gamma _{N_{2}}$ is occurred in all
our developments [15]. Let us apply it to absorption estimation in
Q-branches (Figs. 4, 5). The absorption contour for the pressure $p_{He}=49.6
$ atm is adjusted by the fitting of the halfwidth (Fig. 4) and the value $%
\gamma _{He}=0.35\gamma _{N_{2}}$ has been obtained. Then, the critical
pressure $\gamma _{He}=37.2$ atm has been predicted. Indeed, the
calculations at this pressure and with $\gamma _{He}=0.52\gamma _{N_{2}}$
approximately yield the double absorption in comparison with the Lorentzian,
if the contribution of the P- and R-branches is not taken into account. The
parameter $c$ is equal to 1.2 for the Q-branch, also $a=0.6$ and $b=8$. In
order to compare effects in different branches, their values are chosen to
be close to those used for computations in other bands (e.g., Fig. 2) and it
is clear that their adjustment could improve the spectrum retrieval,
although the calculations in the Q-branch are weakly sensitive to the
parameter $b$. 

\begin{figure}
\centerline{\psfig{figure=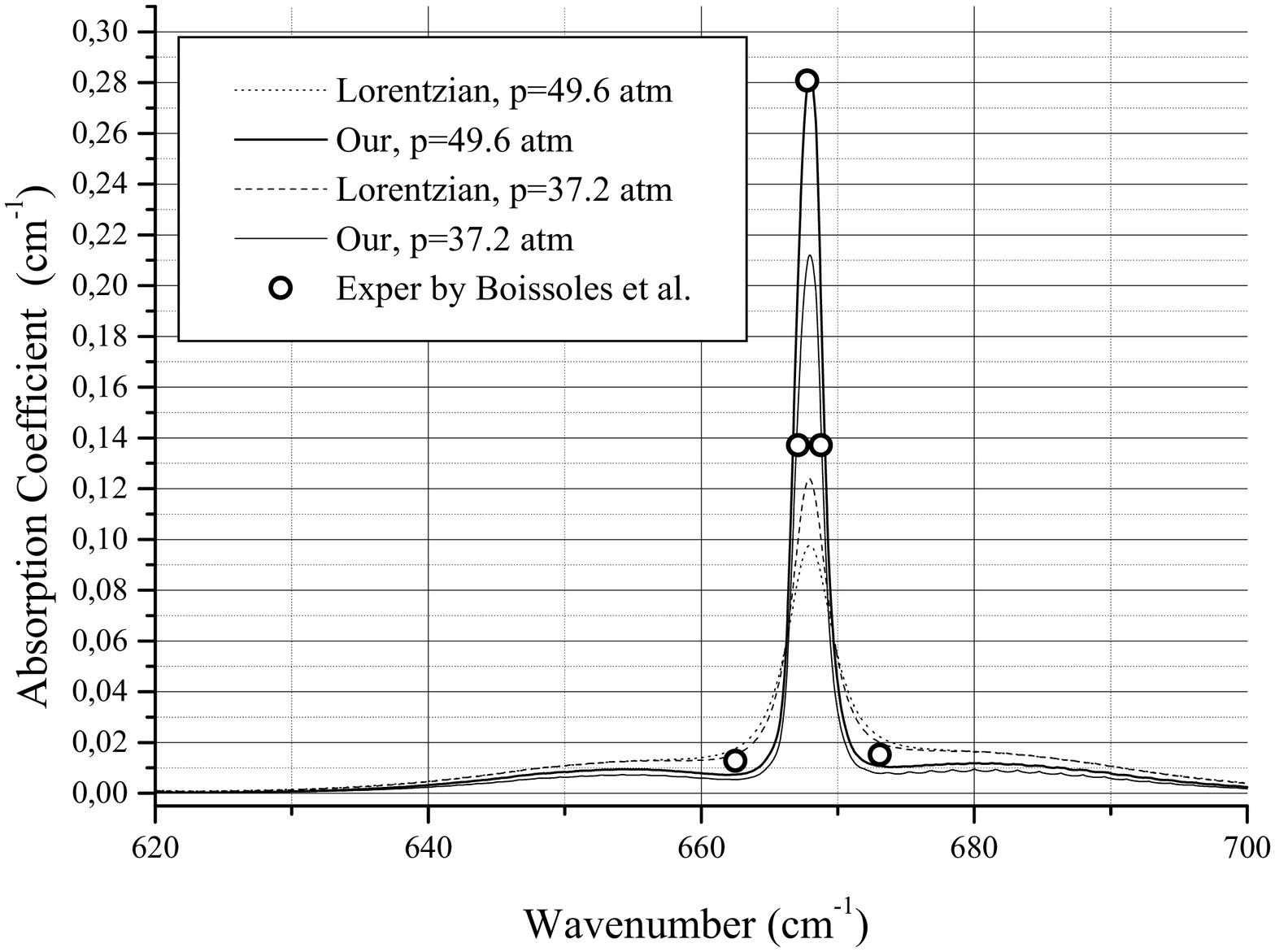,width=8cm,height=5.3cm,angle=270}}
\caption{Application of the main hypothesis of the minimum time limit of the
angular momentum transfer to calculations of the absorption coefficient in
the central region of the band $\nu _{2}$ near 14 $\mu $m (620-700 cm$^{-1}$%
) in mixtures with helium (experiment by Boissoles et al. [10]). The
pressure $p_{CO2}$ is 4.2 Torr. The pressures of He $p_{He}$ is 49.6 atm.
The temperature is 296 K. The double absorption in Q-branch without
contributions of P- and R-branches in comparison with the Lorentzian is
evaluated for the pressure of He equal 37.2 atm. The halfwidth $\gamma_{He} $
of the CO$_{2}$-He absorption lines is $0.52\gamma _{N2}$. The path length
is 3.85 cm.}
\end{figure}

\begin{figure}
\centerline{\psfig{figure=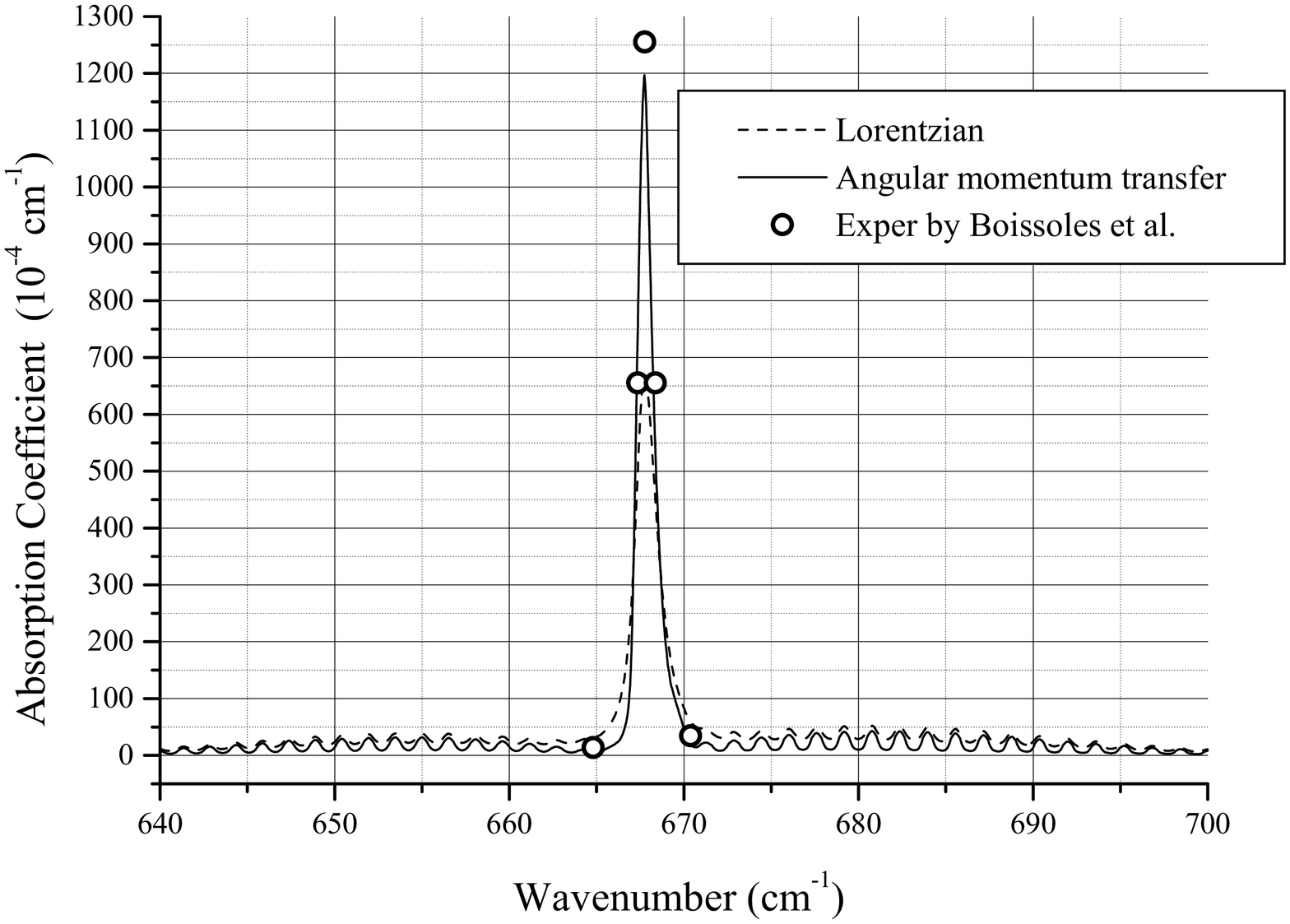,width=8cm,height=5.3cm,angle=270}}
\caption{The attempt to calculate the absorption coefficient in the central
region of the band $\nu _{2}$ near 14 $\mu $m (640-700 cm$^{-1}$, Q-branch)
in mixtures with helium (experiment by Boissoles et al. [10]). The pressure $%
p_{CO2}$ is 1.0 Torr. The pressure of He $p_{He}$ is 9.85 atm. The
temperature is 296 K. The double absorption in comparison with the
Lorentzian is obtained only for the halfwidth of the CO$_{2}$-He absorption
lines $\gamma _{He}=0.64\gamma _{N2}$ (compare with Fig. 4). The path length
is 3.85 cm.}
\end{figure}

According to Ref. [10], the double absorption becomes at pressure $%
p_{He}=9.85$ atm (Fig. 5). However, the parameter $\gamma _{He}$ equals to $%
0.52\gamma _{N_{2}}$ and we do not enable to obtain the double absorption
(Fig. 4) with the parameters defined in Fig. 3. The strong line by line
analysis of the halfwidth values has been accomplished in Ref [10]. Our
testing of the Lorentzian contribution for many samples in Ref. [10] show
that these values [10] slightly differ ($\gamma _{He}\approx \gamma _{N_{2}}$%
) from those stored in databases [28, 29]. These discrepancies are different
[10] for various branches and spectral regions. In any case, they are
greater than our value $\gamma _{He}=0.52\gamma _{N_{2}}$. In turn, our
value $0.52\gamma _{N_{2}}$ successfully applied to many regions is near to
the experimental value of $0.59\gamma _{N_{2}}$ by Burch et al. [2].
Certainly, the conventional cutoff, e.g., no wings greater than 600 cm$^{-1}$%
, is used in all calculations with the Lorentzian. The narrowing problem
based on the Lorentzian priority shows in Fig. 5 that evaluation of the
Lorentz's halfwidth is significant and it is deserved for detailed studies.
Hitherto the rough relationship $\gamma _{He}=0.52\gamma _{N_{2}}$ suffices
to describe main features of the effect of interest. The spectroscopic
information stored in the database GEISA-97 [28] (see also Ref. [29]) is
used in all computations.

\subsection{3$\nu _{3}$ band region}

Computations in the 3$\nu _{3}$ region are presented in Fig. 6. The pressure
of helium (Fig. 6a) is approximately the same as in Fig. 2a and it may be
associated with the critical pressure for the P-branch. Thus, the absorption
must be reckoned with the $\gamma _{He}=0.52\gamma _{N_{2}}$ in the P-branch
region. We also found that for the R-branch that contains a strong head near
the branch center this parameter is $0.2\gamma _{N_{2}}$. This means that
the Lorentzian halfwidths for lines in R-branch reaches its saturation
values at pressures in $0.2/0.52$ times lesser than for lines of P-branch.
Thus, there are two intervals in the 3$\nu _{3}$ region that may be
described with the different parameters. The parameter $c$ is 1.6 and 1.2
for the P- and R-branches, respectively. The parameter $b$ is the same for
two branches, $b=8$. The absorption modeling in the P-branch may be
considered as being extrapolated from other regions, but the absorption in
the head of the R-branch is fitted. The extrapolation is demonstrated in
Fig. 6b for the higher pressures.

\begin{figure}
\centerline{\psfig{figure=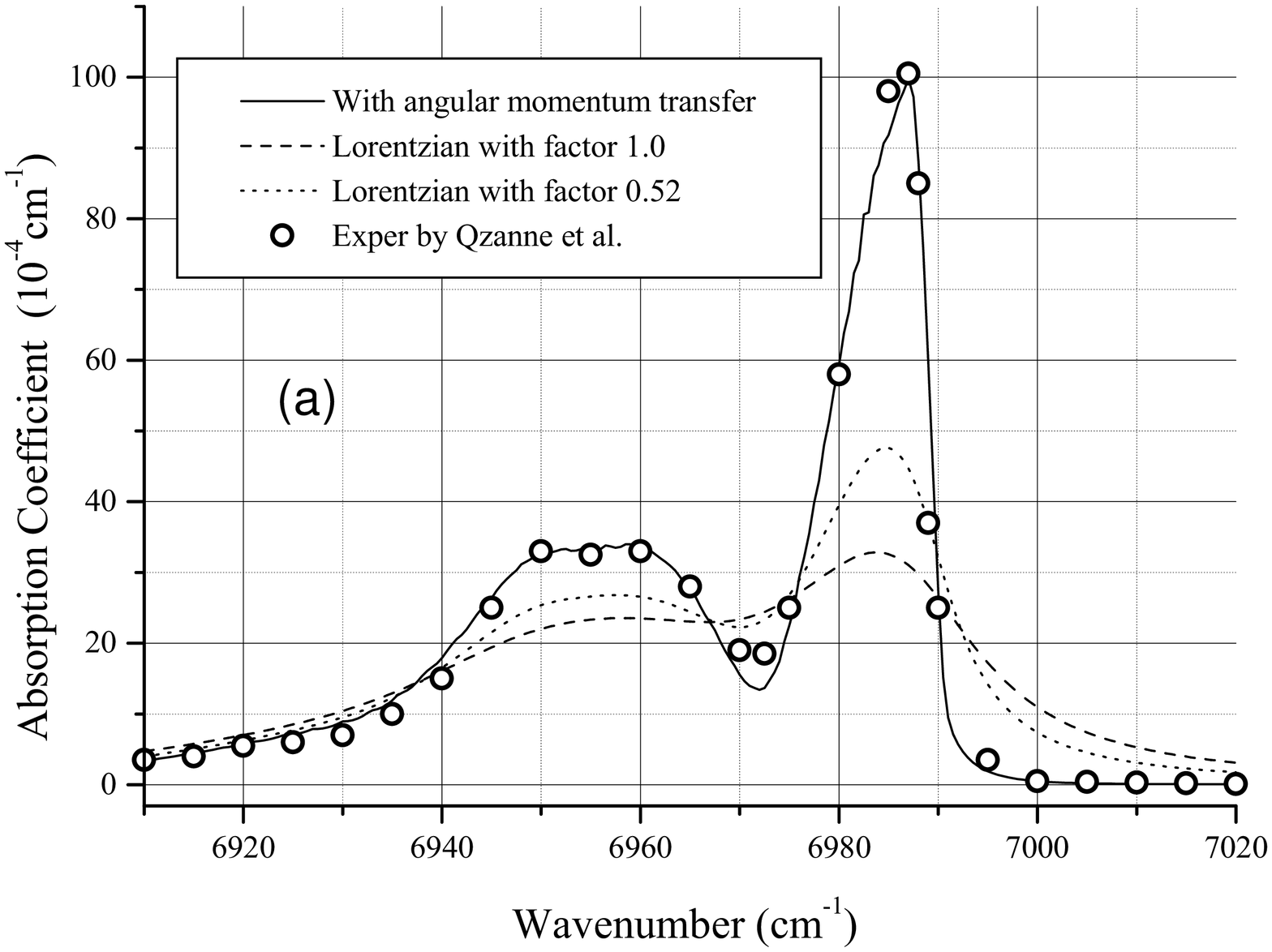,width=8cm,height=5.3cm,angle=270}}
\centerline{\psfig{figure=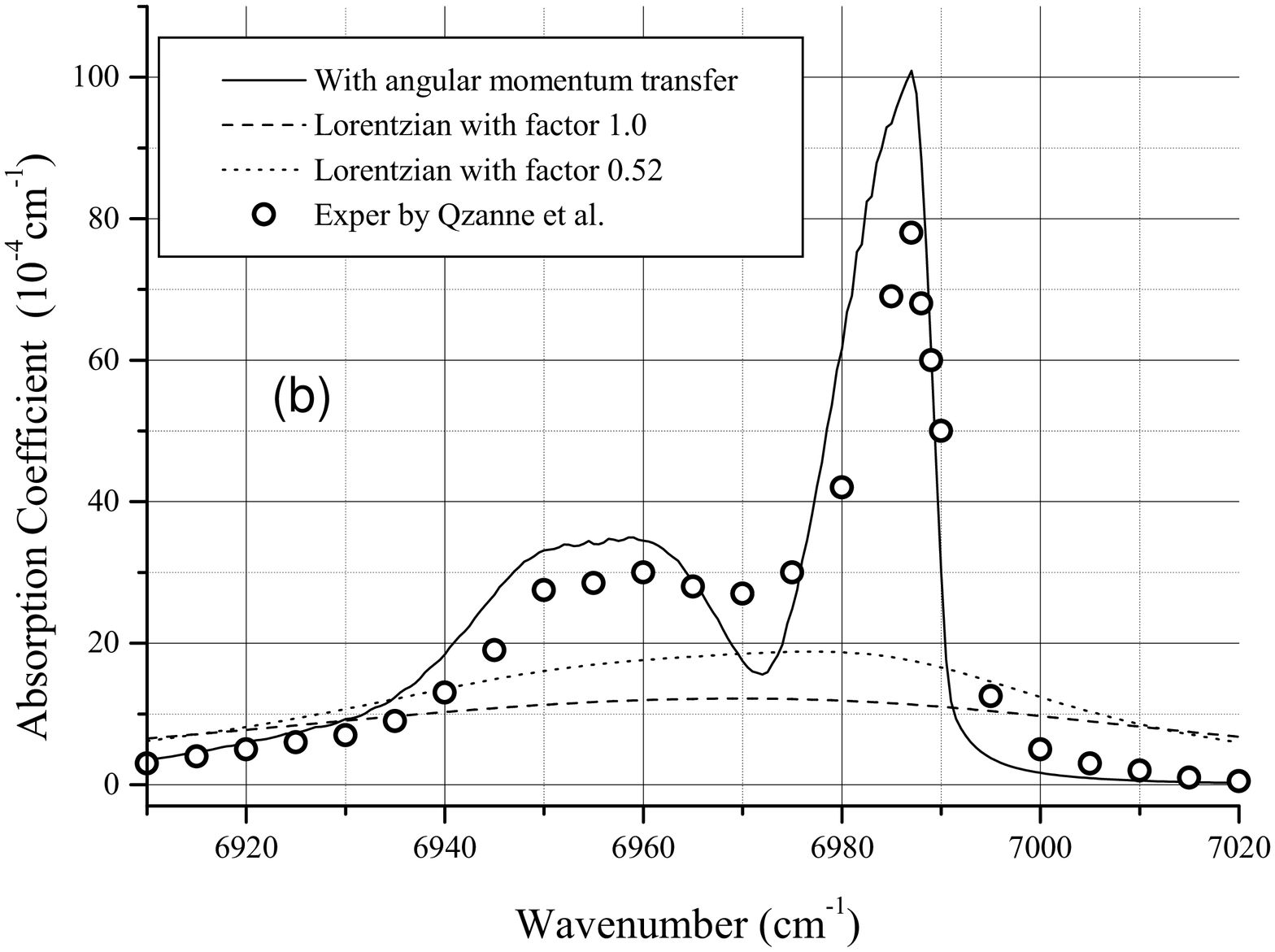,width=8cm,height=5.3cm,angle=270}}
\caption{Verification of the main hypothesis of the minimum time limit of
the angular momentum transfer on the example of the two calculations of the
absorption coefficient in the central region of the band $\nu _{3}$ near 1.4 
$\mu $m (P- and R-branches with the head) in mixtures with helium.
Densities: (a) $n_{CO2}=4.62$ Am and $n_{He}=121.2$ Am (131.86 atm); (b) $%
n_{CO2}=4.66$ Am and $n_{He}=598.7$ Am (645.41 atm). Experiment by Ozanne et
al. [9] was performed at the temperature 297 K.}
\end{figure}

Contrary to data in Fig. 2b, the discrepancies in Fig. 6b have other
character. The effect corresponds to the branch-broadening not the
branch-narrowing as in Fig. 2b. Probably, the reorientation of the angular
momentum of molecules, that concerns the inter-branch interaction, can lead
to decrease of the period of the angular momentum transfer from photon to
molecules. A trivial explanation also is not excluded that there is other
set of the adjusting parameters, that allows us to obtain a description
similar to that in Fig. 2b. A picture similar to the small inter-branch
narrowing effect is noticeable in Fig. 6a.

\section{SUMMARY AND CONCLUSION}

The present studies and modelings followed confirm the main hypothesis of
this communication that the Lorentzian halfwidth of absorption lines trends
toward approaching to the saturation value at high pressures. Also, the
lower time limit of the angular momentum transfer seems as an encouraged
hypothesis with the reasonable physical meaning. We tried to follow the
Lorentzian priority that, in our opinion, underlines the wave properties of
the phenomena considered. However, as our modelings of absorption in the CO$%
_{2}$-He mixture show, the simple redistribution of the intensity in the
band with the direct construction of the Lorentzian halfwidth $\gamma $,
e.g., in the form 
\begin{equation}
\gamma =\frac{\gamma _{c}\gamma _{rot}^{s}}{\gamma _{c}+\gamma _{rot}^{s}}
\end{equation}
does not properly describe the needed narrowing effect. Although this evident 
formula seems to produces the needed double absorption in the band center,
for the halfwidth $\gamma $ could be equals to $\gamma _{c}/2$ at the
critical pressure $p_{s}$, when $\gamma _{c}=\gamma _{rot}^{s}$,
nevertheless the narrowing function (4.1) must be introduced for these aims.
The halfwidth in Eq. (6.1) is equal to the conventional halfwidth $\gamma
_{c}$ at small pressures and to the saturated halfwidth $\gamma _{rot}^{s}$
at high ones.

The Lorentzian priority is seen from the formula for ''experimental''
absorption coefficient that has been constructed in the following form [16] 
\begin{eqnarray}
\alpha _{eff}\left( \omega \right) &=&\alpha \left( \omega \right) + \\
&&\frac{1}{2x}\ln \left( 1+\frac{b\left( \omega \right) }{2\alpha \left(
\omega \right) }\left( 1-\exp \left( -2\alpha \left( \omega \right) x\right)
\right) \right) .\text{ }  \nonumber
\end{eqnarray}
that comprises two terms connected with linear and nonlinear absorption. The
nonlinear absorption is large in water vapor, but it is small in mixtures of
CO$_{2}$ with various broadeners, therefore the problem of the line shape
may be first of all regarded as the sub-Lorentzian behavior of the linear
absorption $\alpha \left( \omega \right) $ that is determined by the {\it %
line by line} summing over $n$ lines, 
\begin{eqnarray}
\alpha \left( \omega \right) &=&\sum\limits_{i}^{n}N\left( p,\omega
_{i},T\right) FL\left( \omega ,\omega _{i},T\right) \times \\
&&Ab\left( \omega ,\omega _{i},T,M=5\right) \Gamma \left( \omega ,\omega
_{i}\right) .  \nonumber
\end{eqnarray}
The conventional factor $N\left( p,\omega _{i},T\right) $ is the molecule
number in the lower state of the transition $i$ at the temperature $T$. This
factor contains the Gibbs (Boltzmann) exponent. The full Lorentz line shape $%
FL\left( \omega ,\omega _{i},T\right) $ includes the integral intensity of
the line at the temperature $T$ per one molecule. The universal narrowing
function $\Gamma \left( \omega ,\omega _{i}\right) $ is presented by Eq.
(4.1) for high pressures. The Fermi (anti-Fermi) absorption factor $Ab\left(
\omega ,\omega _{i},T,M\right) $ for the right (left) wing is taken with the 
$M=5$. The parameter $M$ manifests the slope of the Fermi-distribution that
presents an exponential form of the absorption in far wings. Since the
exponential form is not essential in the resonance region, this factor as
well as nonlinear absorption with the nonlinear absorption coefficient $%
b\left( \omega \right) $ is not considered in this communication. The
Lorentzian priority is presented in Eq. (6.3) by the fact that the full
Lorentz's line shape enters Eq. (6.3) as a single factor.

\begin{figure}
\centerline{\psfig{figure=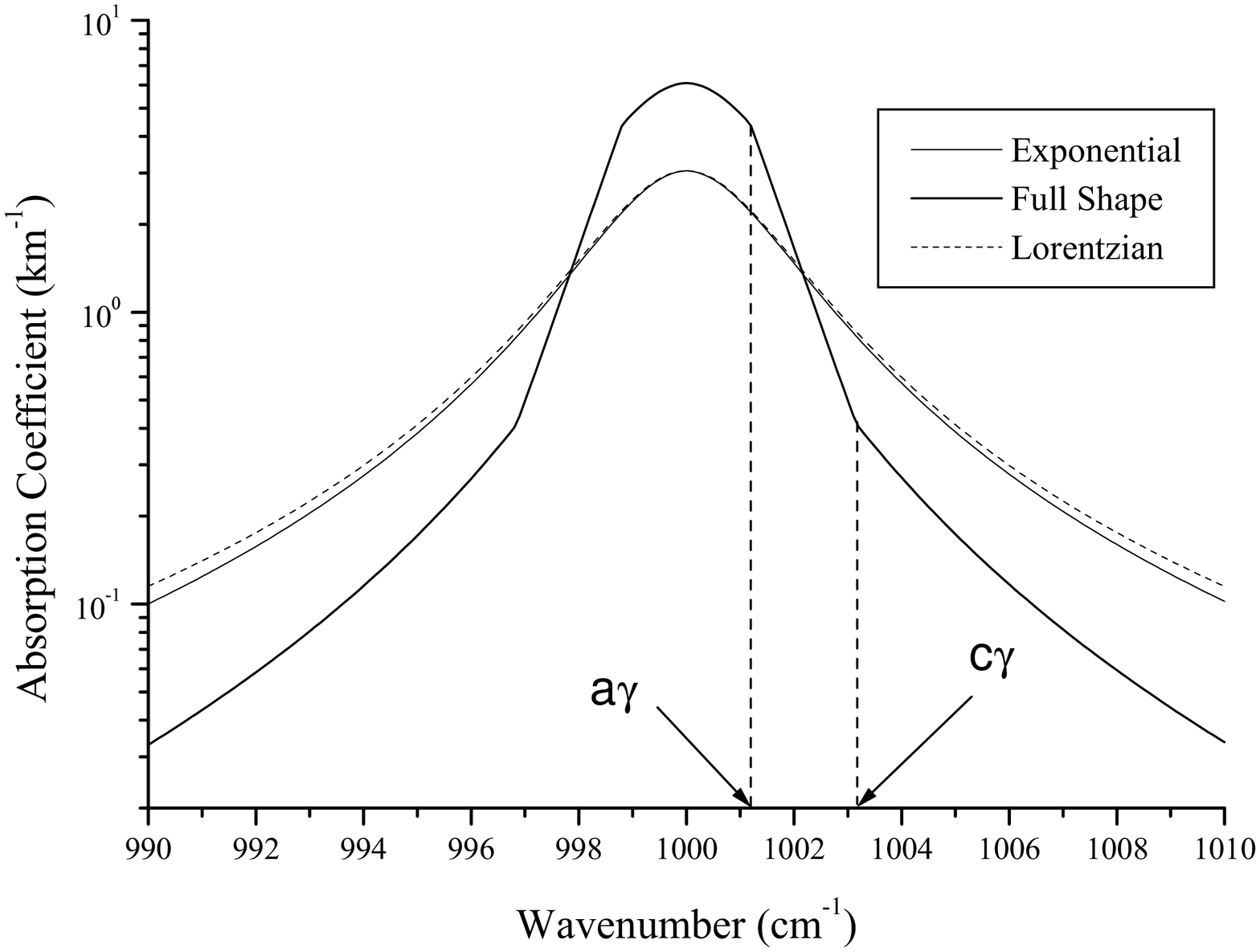,width=8cm,height=5.3cm,angle=270}}
\caption{The high pressure hypothetical line shape of a single line
positioned at 1000 cm$^{-1}$ in the logarithmic scale has the simple form.
The Lorentzian is the full Lorentz line shape. The exponential form is the
Lorentzian with some Fermi or anti-Fermi factor. The full line shape is the
Lorentzian with the narrowing function (4.1) and Fermi or anti-Fermi factor.
The nonlinear absorption due to the specific line mixing (self-mixing for
the single line) [16] that leads to increase of the absorption in near wings
is not presented.}
\end{figure}

The line shape of the single line for high pressures is depicted in Fig. 7.
The rapid decrease of its value near the value $c\gamma _{rot}^{s}$ that
associates with the separation of the rotational line manifests the angular
momentum transfer in the photon absorption act, i.e., due to the inertia
momentum of molecules. Adding the nonlinear absorption increases the
absorption in near wings and the non-Lorentzian line shape in Fig. 7 with
this enlargement in the central region can resemble that what was called as
the ''generalized line shape'' [31,32] (see also Ref. [33]). This is not
surprising because the analysis of the experimental facts in far wings was
the same as in our works as in Refs. [31-33]. Nevertheless, the methods and
the explanation of these phenomena in these references are different from
ours.

We try to utilize facts and observations from Refs. [11,12] illustrated the
common absorption features inherited, in our opinion, to the interaction of
the electromagnetic field with a two-level system and independent from
intermolecular forces. As a result, the very simple formulas (6.2) and (6.3)
allow one to provide the prompt computations of CO$_{2}$ absorption in the
total spectrum region for various broadeners [15]. Only the exponential form
somewhat enlarges the chip time. As it is shown in Fig. 7, this form may be
ignored for computations in the central region, e.g., in the Q-branches. In
general, this approach should be useful, first of all, for atmospheric
applications. Further parametrization of the narrowing function must be
fulfilled for moderate pressures. This means that the detailed dependence of
the halfwidth on pressure (Fig. 1) must be stipulated.

The studies of this paper help us to exhibit main problems of the line shape
in the closed logical scheme as follows. The line shape is broadened in the
electromagnetic fields, so that the absorption coefficient $k$ is described
by the relation $k\sim 1/\left( \delta \omega ^{2}+\gamma ^{2}+\eta I\right) 
$, where $\delta \omega $ is the detuning wavenumber, $\gamma $ is the
Lorentz's halfwidth, $I$ is the radiation intensity with the proportionality
constant $\eta $. This relationship gets an origin from Ref. [34] and it
presents the nonlinear absorption near the resonance region [25]. It is
conventionally deemed as a formula that cannot describe the absorption in
far wings. However, the expansion of this expression for far wings over the
small third term in the denominator at small intensities can lead to the
nonlinear absorption equation [16,17], if this expression is substituted for
the $k$ in the Beer-Lambert law and the term of the second power of $I$ is
omitted. The latter can be done, in our opinion, according to the causality
principle [16,17]. Some observed features of the far wing absorption had
been obtained [16,17,15] with the help of this nonlinear absorption. Thus,
the nonlinear absorption as in the resonance region [34,25] as in the far
wings [16,17] should have the same origin.

Also, the same technique for deriving the nonlinear constants should be
occurred in both these regions. The rise of the so-called ''collapse'' of
the wave function, the main paradox of the quantum mechanics [36], had been
excluded in Refs. [16,17] via its compensation by the absorption (chaos).
The appearance of the photon fluctuations associated with two-level systems
[15] in the resonance region for the period, when the system changes
stationary states, may be also associated with this ''collapse'' of two
coherent photons and the increase of the absorption in two times in the
resonance region may be formally considered in a linear theory as the
compensation of this ''collapse'' effect.

The present remarks into the line shape theory should be attractive for a
few reasons. Firstly, it agrees with the local photon fluctuation of five
photons associated with a two-level system (used in Ref.[15] for far wing
calculations). Secondly, there is a hope that old 
Einstein's hypothesis of the induced emission in the two-level system
that requires consideration, at least, two photons in one act of the
absorption/emission should be properly interpreted. Also, the problems of
the interaction of radiation with matter, not only intermolecular
interactions, have been highlighted as important in the line shape theory.
The latter underlines a close connection of the absorption as in resonance
regions as in wings with the classical impact theory by Lorentz, Lenz, and
Weisskopf [19]. Still one question should be interesting. How means the
technique of the density matrix could incorporate the local photon
fluctuations or some radical changes should be required, if the present
ideas will be admitted in the line shape theory?

\acknowledgments

The author would be grateful to colleagues: S.N.Mikhailenko for processing
some artificial synthetic spectra that helped us to select ideas on initial
stages of this work, V.I.Perevalov for many consultations over the
spectroscopic information for CO$_{2}$ molecule, A.V.Nikitin and A.A.Chursin
for the help in the decision of some technical problems, A.A.Chursin and
S.A.Tashkun for providing with copies of a few unavailable for us papers.
Also, Dr. V.P.Kochanov is appreciated for fruitful discussions of the
nonlinear absorption. 
\[
\]

Email: gvf@lts.iao.ru


\begin{references}
\bibitem{}  B.H. Winters, S. Silverman, and W.S. Benedict, J. Quant.
Spectrosc. Radiat. Transfer, {\bf 4}, 527 (1964).

\bibitem{}  D.E.Burch, D.A. Gryvnak, R.R. Patty, and C.E. Bartry, J. Opt.
Soc. Am., {\bf 59}, 267 (1969).

\bibitem{}  N.Bloembergen, E.M.Purcell, and R.V.Pound, Phys. Rev., {\bf 73},
679 (1948).

\bibitem{}  A.L\'{e}vy, N.Lacome, and C.Chackerian, Jr., in {\it The
Spectroscopy of the Earth's Atmosphere and Interstellar Medium}, edited by
K.Narahari Rao and A.Weber, Academic Press Inc., San Diego, 261 (1992).

\bibitem{}  D.Sp\"{a}nkuch, Atm. Res., {\bf 23}, 323 (1989).

\bibitem{}  I.M.Grigor'ev, V.M.Tarabukhin, M.V.Tonkov, Opt. spectrosk., {\bf %
58}, 244 (1985) (in Russian).

\bibitem{}  F.Thibault, J.Boissoles, R. Le Doucen, V. Menoux, and C. Boulet,
J.Chem. Phys., {\bf 100}, 210 (1994).

\bibitem{}  J.Boissoles, F.Thibault, R. Le Doucen, V. Menoux, and C. Boulet,
J.Chem. Phys, {\bf 100}, 215 (1994).

\bibitem{}  L. Ozanne, Nguyen-Van-Thanh, C. Brodbeck, J.P.Bouanich, J.-M.
Hartmann, and C. Boulet, J.Chem. Phys., {\bf 102}, 7306 (1995).

\bibitem{}  J.Boissoles, F.Thibault, and C. Boulet, J. Quant. Spectrosc.
Radiat. Transfer, {\bf 56}, 835 (1996).

\bibitem{}  L.L. Strow and D.Reuter, Appl. Opt., {\bf 27}, 872 (1988).

\bibitem{}  R.Rodrigues, K.W.Jucks, N.Lacome, Gh.Blanquet, J.Walrand,
W.A.Traub, B.Khalil, R. Le Doucen, A.Valentin, C.Camy-Peyret, L.Bonamy, and
J.-M.Hartmann, J. Quant. Spectrosc. Radiat. Transfer, {\bf 61}, 153 (1999).

\bibitem{}  C.P.Rinsland and L.L.Strow, Appl. Opt., {\bf 28}, 457 (1989).

\bibitem{}  D.P.Edwards and L.L. Strow, J. Geophys. Res., {\bf 96}, 20859
(1991).

\bibitem{}  V.F.Golovko, in {\it Proc. IRS-2000. Current Problems in
Atmospheric Radiation}, A.DEEPAK Publishing, 2001.

\bibitem{}  V.F.Golovko, J. Quant. Spectrosc. Radiat. Transfer, {\bf 65},
621 (2000).

\bibitem{}  V.F.Golovko, J. Quant. Spectrosc. Radiat. Transfer, {\bf 65},
821 (2000).

\bibitem{}  P.W.Rosenkranz, J. Chem. Phys., {\bf 83}, 6139 (1985).

\bibitem{}  See, e.g., R.G.Breene, Jr., {\it The Shift and Shape of Spectral
Lines} (Pergamon Press, Oxford, 1961).

\bibitem{}  See, e.g., H.Margenau and M.Lewis, Rev. Mod. Phys., {\bf 31},
569 (1959).

\bibitem{}  E.W.Smith, V.Giraud, and J.Cooper, J. Chem. Phys., {\bf 65},
1256 (1976).

\bibitem{}  See, e.g., C. Boulet, D.Robert, and L.Galatry, J. Chem. Phys., 
{\bf 72}, 751 (1980).

\bibitem{}  S.A.Clough, F.X.Kneizys, and R.W.Davies, Atm. Res. {\bf 23}, 229
(1989).

\bibitem{}  J.R. Murray and A. Javan, J.Mol. Spectrosc., {\bf 42}, 1 (1972).

\bibitem{}  See, e.g., S.Stenholm, {\it Foundation of Laser Spectroscopy} (A
Wiley-Interscience Publication, John Wiley and Sons, N.Y., etc, 1984); (Mir,
Moscow, 1987) (in Russian). V.S.Letokhov and V.P.Chebotaev, {\it Nonlinear
laser high-resolution spectroscopy} (Nauka, Fizmatgiz, Moscow, 1990) (in
Russian).

\bibitem{}  U.Fano, Phys. Rev., {\bf 131}, 259 (1963).

\bibitem{}  Yu.S.Makushkin, A.I.Petrova, Izv. Vyssh. Ucheb. Zaved., Fiz., 
{\bf No. 12}, 99 (1986). For the band 3$\nu _{3},$Yu.S.Makushkin,
S.D.Tvorogov, L.I.Nesmelova, and A.I.Petrova, ''Studies of influence of
spectral exchange on formation of the line shape of vibration-rotational
transitions in CO$_{2}$'', VIIIth All-Union Symposium on High-Resolution
Spectroscopy, (Krasnoyarsk, June 1987, Russia), unpublished.

\bibitem{}  N. Jacquinet-Husson et al., J. Quant. Spectrosc. Radiat.
Transfer, {\bf 62}, 205 (1999).

\bibitem{}  L.S.Rothman et al., J. Quant. Spectrosc. Radiat. Transfer, {\bf %
60}, 665 (1998).

\bibitem{}  V.F.Golovko, J. Quant. Spectrosc. Radiat. Transfer, {\bf 69},
431 (2001).

\bibitem{}  G.V.Telegin snd V.V.Fomin, preprint No. 26 (Institute of
Atmospheric Optics SB RAS, Tomsk, 1979) (in Russian).

\bibitem{}  V.V.Fomin, {\it Molecular Absorption in Infrared Transparency
Windows} (Nauka, Siberian division, Novosibirsk, 1986) (in Russian).

\bibitem{}  L.I.Nesmelova, O.B.Rodimova, and S.D.Tvorogov, {\it Spectral
Line Shape and Intermolecular Interaction} (Nauka, Siberian division,
Novosibirsk, 1986) (in Russian).

\bibitem{}  R.Karplus, I.A.Schwinger, Phys. Rev., {\bf 73}, 1020 (1948).

\bibitem{}  P.A.M.Dirac, {\it The principles of quantum mechanics}, 4th ed.
(at the Clarendon Press, Oxford, 1958) Sec. ''Photon interference''; 2nd ed.
(Fizmatgiz, Moscow, 1979), p. 21 (in Russian).
\end{references}
\end{document}